\documentclass[12pt,a4paper,american]{revtex4}
\usepackage[T1]{fontenc}
\usepackage[latin1]{inputenc}
\usepackage{amsmath}
\usepackage{graphicx}
\usepackage{amssymb}

\usepackage{babel}

\begin{document}

\title{Creation of massive particles in a tunneling universe}

\author{Jooyoo Hong$^{1,2}$, Alexander Vilenkin$^1$ and Serge Winitzki$^3$}

\affiliation{$^1$ Institute of Cosmology, Department of Physics and
Astronomy, Tufts University, Medford, MA 02155, USA}

\affiliation{$^2$ Department of Physics, Hanyang University at Ansan,
Ansan, Kyunggi-do, 425-791 Korea}

\affiliation{$^3$ Department of Physics, Ludwig-Maximilians University,
80333 Munich, Germany}

\begin{abstract}
We examine the particle production during tunneling in quantum cosmology.
We consider a minisuperspace model with a massive, conformally coupled
scalar field and a uniform radiation background.
In this model, we construct a semiclassical wave function describing a 
small recollapsing universe and a nucleated inflating universe 
(``tunneling from something'').  We find that the quantum states of the 
scalar field in both the initial and the nucleated universe are close to 
the adiabatic vacuum, the number of created particles is
small, and their backreaction on the metric is negligible. We show that
the use of the semiclassical approximation is justified for this
wave function.
Our results imply that the creation of the universe from
nothing can be understood as a limit of tunneling from a small
recollapsing universe.
\end{abstract}
\maketitle
\section{Introduction}

A semiclassical picture of quantum cosmology based on the
Wheeler-DeWitt equation describes tunneling from a state of vanishing
size (``tunneling from nothing'') to a closed inflating universe. The
process of tunneling from nothing can be thought of as a limit of
tunneling from a closed recollapsing universe of very small but
nonzero size to an inflating universe (``tunneling from
something''). This paper continues the investigation of particle
creation during tunneling in quantum cosmology. Conflicting claims of
excessive particle production that invalidates the semiclassical
approximation \cite{R84,LRR02}, on the one hand, and of essentially no
particle content in the nucleated universe \cite{V88,VV88,GV97}, on
the other hand, have been advanced in the literature. Our intent is to
resolve this long-standing controversy.

The process of tunneling from a recollapsing universe sensitively
depends on the quantum state of that universe. In the companion paper
\cite{JVW1} we have shown that, at least in the case of a massless
field, the results of Rubakov \emph{et~al.}~\cite{R84,LRR02} should be
interpreted not as an indication of a large particle production but as a
consequence of an inadequate choice of the initial quantum state of
the universe. A generic quantum state of the recollapsing universe
will contain a superposition of various semiclassical geometries.
Some geometries in the superposition do not describe a nucleated
universe but rather a universe that expanded from zero size
``over the barrier'' without
tunneling, because it contained a large number of particles. Other
geometries in the superposition will describe a nucleated universe
with a small number of particles. One can hope to extract well-defined
particle numbers from a Wheeler-DeWitt wave function only if it
describes a single semiclassical background spacetime. However, the
process of tunneling amplifies the differences between branches of the
wave function.
A semiclassical branch of the wave function describing
a universe with a large particle content may be strongly
suppressed in the initial recollapsing universe but may dominate the
wave function of the tunneling universe (because it avoids being 
exponentially suppressed during tunneling).

For a massless field, there is a well-defined vacuum and the particle
production is absent \cite{Parker}.  The quantum state chosen by
Rubakov \cite{R84} corresponds to a squeezed state with respect to that
vacuum and consists of an infinite superposition of excited states with
all possible energies.  Each excited state
is described by a branch of the wave function that has its own semiclassical
geometry. As we showed in Ref.~\cite{JVW1}, one cannot neglect the
difference between these semiclassical geometries if one considers the
nucleated universe. We call this phenomenon
``critical branching''. We have shown that the ``catastrophic
particle creation'' found by Rubakov \emph{et~al.}~is in fact a sign of
critical branching that happens with their choice of the initial state.
If one chooses the initial
state to be the vacuum, or any excited state that is an eigenstate of
the particle occupation numbers, then there is no branching,
and the wave function describes a single semiclassical geometry.  

The purpose of the present paper is to extend the analysis
of Ref.~\cite{JVW1} to the case of a massive field.
A nonzero mass $m$ introduces important qualitative differences into
the problem.  First, a massive scalar field is nontrivially coupled to
the time-dependent metric, and there is necessarily some particle
production. Even if one imagines starting the recollapsing universe in
a vacuum state at early times, the field will be in a superposition of
states with all possible occupation numbers at other times. Additional
creation of excited states may occur as a result of tunneling.
Thus, it is not clear \emph{a priori} that critical branching can be
avoided for any choice of the initial state.  Moreover, the definition
of the vacuum state for a massive field in an expanding universe is
notoriously ambiguous; as a result, the number of particles is also
subject to ambiguity.
In this paper we give a detailed analysis of these issues.

The approach taken by Rubakov \emph{et~al.}~\cite{R84,LRR02} was to
define the vacuum state by diagonalizing the scalar field Hamiltonian
at a moment of time.  However, this procedure  is problematic since it
is known to give unphysical results for the particle density in some
cases \cite{Fulling}.  A well-motivated definition of vacuum in a
slowly changing background geometry has been developed by Parker and
Fulling; it is the so-called adiabatic vacuum~\cite{Parker,PF74}.  An
adiabatic vacuum of $k$-th order is defined using a truncated
asymptotic series obtained from the WKB expansion.  It has some
dependence on the moment of time when it is defined and on the order
$k$ at which the series is truncated.  The resulting uncertainty is of
order of the $k$-th power of the adiabatic parameter.  On the other
hand, it follows from our analysis in \cite{JVW1} that an admissible
state of the universe must be defined with an exponential accuracy to
avoid the ``critical branching''. We need to specify a vacuum state
with a higher accuracy than the definition of the adiabatic vacuum
allows.

Our main result in this paper is a prescription for constructing a
well-behaved semiclassical state of the universe which does not exhibit
critical branching.  We show that this state always exists and
coincides with the adiabatic vacuum within the accuracy to which the
latter can be defined. It also coincides with the canonically defined
vacuum in the massless case. 

The paper is organized as follows. In
Sec.~\ref{sec:The-adiabatic-vacuum} we construct a family of
well-behaved semiclassical wave functions of the universe.
We give a prescription to
select a unique quantum state that we call the ``Gaussian vacuum''. In
this state the backreaction of produced particles is negligible and
there is no branching. We obtain approximate expressions for the wave
function of this state. In Sec.~\ref{sec:Interpretation} we give a
particle interpretation of the Gaussian vacuum state. We show that the
Gaussian vacuum is indistinguishable from an adiabatic vacuum state to
all allowed orders of the adiabatic expansion. In
Sec.~\ref{sec:Discussion} we discuss our results as well as the
issues raised by Rubakov \emph{et~al.}
We also compare our results to those of Bouhmadi-López
\emph{et~al.}~\cite{BGG02} who addressed some of the same issues.
In the Appendices we give
technical details of the calculations and, in particular, check the
validity of our approximations.

\section{The Gaussian vacuum state}

\label{sec:The-adiabatic-vacuum}

We consider a homogeneous closed FRW metric and an inhomogeneous massive
conformally coupled scalar field. The classical action of the system
is
\begin{eqnarray}
S & = & \int d^4x\sqrt{-g}\left\{ \frac{R}{16\pi }
-\frac{3}{8\pi } H^2\right.\nonumber \\
&  & \left.+\frac{1}{2}\left(\partial _{\mu }\phi
\right)^2-\frac{1}{2}m^2\phi ^2-\frac{1}{12}R\phi ^2\right\} .
\end{eqnarray}
Here $R$ is the scalar curvature and the parameter $H$ represents
the cosmological constant. We use the Planck units, $G=\hbar =c=1$.
We assume $H\ll 1$ and  $m\ll 1$ as a typical physically motivated case.

We expand the inhomogeneous scalar field in spherical harmonics on
the 3-sphere,\begin{equation}
\phi (x,t)=\frac{\pi \sqrt{2}}{a(t)}\sum _{n,l,p}\chi
_{nlp}(t)Q_{lp}^n\left({\mathbf
x}\right).\label{eq:field-chi}\end{equation} A rescaling by the factor
$a(t)$ is done for convenience. Below, only the index $n$ will enter the
equations, and we suppress the indices $l$, $p$ of the modes $\chi
_{nlp}(t)$. The summation over degenerate indices $l$, $p$ spans $l=0$,
..., $n-1$ and $p=-l$, ..., $l$ and introduces an extra factor $n^2$
which we shall insert in explicit calculations below.

The wave function of the universe depends on all modes $\chi _n$
of the scalar field, $\Psi =\Psi \left(a,\left\{ \chi _n\right\} \right)$.
The Wheeler-DeWitt equation, after appropriate rescalings \cite{V88},
takes the form \begin{eqnarray}
\left[\hbar ^2\frac{\partial ^2}{\partial a^2}-a^2+H^2a^4\right]\Psi  &  &
\nonumber \\ +\sum _n\left[-\hbar ^2\frac{\partial ^2}{\partial \chi
_n^2}+n^2\chi _n^2+m^2a^2\chi _n^2\right]\Psi  & = & 0.
\end{eqnarray}
Here and below we explicitly write the Planck constant $\hbar \equiv 1$
only as a formal bookkeeping parameter, to clarify the use of the
WKB approximation.

In addition to the scalar field $\phi $ we now include a small amount
of homogeneous radiation with energy density\begin{equation}
\rho _r=a^{-4}\varepsilon _r,\end{equation}
where $\varepsilon _r>0$ is a constant parameter. Then the Wheeler-DeWitt
equation is modified,\begin{equation}
\left[\frac{\hbar ^2\partial ^2}{\partial a^2}-V(a)+\varepsilon _r+\sum
_n\left[-\frac{\hbar ^2\partial ^2}{\partial \chi _n^2}+\omega _n^2\chi
_n^2\right]\right]\Psi =0,\label{eq:wdw}\end{equation} where we have
defined\begin{eqnarray} V(a) & \equiv  & a^2-H^2a^4,\label{eq:Va}\\
\omega _n(a) & \equiv  & \sqrt{n^2+m^2a^2}.\label{eq:omega-def}
\end{eqnarray}
If restricted to the coordinate $a$, Eq.~(\ref{eq:wdw}) is a stationary
Schrödinger equation for a particle in a potential.

\subsection{Gaussian solutions of the WDW equation}

In the companion paper \cite{JVW1} we have used the method of
Refs.~\cite{BBW73,LR79,HH85,Banks} to find an approximate solution of
Eq.~(\ref{eq:wdw}). The solution may be found as a linear combination of
Gaussian terms of the form\begin{equation} \Psi \left(a,\left\{ \chi
_n\right\} \right)=\exp \left[-\frac{S(a)}{\hbar }-\frac{1}{2\hbar }\sum
_nS_n(a)\chi _n^2\right],\label{eq:Gauss}\end{equation} where $S(a)$ and
$S_n(a)$ are functions to be determined. The functions $S_n(a)$ must
satisfy the condition that we shall call the ``regularity condition'',
\begin{equation}
0<Re\, S_n(a)<+\infty \textrm{ for all }a,\,
n.\label{eq:Sn-pos}\end{equation} With this condition, the Gaussian wave
function of Eq.~(\ref{eq:Gauss}) is well-defined everywhere and quickly
decays at large $\chi _n$. In Ref.~\cite{JVW1} we have shown that a
violation of the regularity condition indicates a splitting of the wave
function into decoherent branches with macroscopically different
semiclassical geometries and different particle contents
(critical branching).
A wave function
of the universe can be consistently interpreted in terms of a classical
spacetime with a quantum field only if a single underlying semiclassical
geometry is present. Each term of the form (\ref{eq:Gauss}) will describe a
single semiclassical geometry if the regularity condition holds for its
function $S_n(a)$. Then the branches of the wave function corresponding
to each such term will describe independent, decoherent semiclassical
universes. Therefore we may impose the regularity condition on all
terms of the form of Eq.~(\ref{eq:Gauss}) that comprise the wave
function of the universe.

We now substitute the ansatz of Eq.~(\ref{eq:Gauss}) into
Eq.~(\ref{eq:wdw}) and obtain\begin{eqnarray}
\left(S^{\prime }\right)^2-V(a)-\hbar S^{\prime \prime }+\hbar \sum _nS_n &
= & 0,\label{eq:S0equ}\\ \left[S^{\prime }S_n^{\prime }-S_n^2+\left[\omega
_n(a)\right]^2-\frac{\hbar }{2}S_n^{\prime \prime }\right]\chi _n^2 &  &
\nonumber \\ +\frac{1}{4}(S_n')^2\chi _n^4 & = & 0.\label{eq:Snequ}
\end{eqnarray}
Our approximation consists of disregarding the terms of order $O(\hbar )$
and $O(\chi _n^4)$; the applicability of this approximation
is analyzed in Appendix~\ref{sec:app-Applicability}. We then have the
following equations for the functions $S(a)$ and $S_n(a)$,\begin{eqnarray}
& (S')^2-V(a)+\varepsilon _r & =0,\label{eq:S0eq1}\\
& S'S_n'-S_n^2+\omega _n^2 & =0.\label{eq:Sneq1}
\end{eqnarray}

\begin{figure}
\begin{center}\includegraphics[  width=3in,
keepaspectratio]{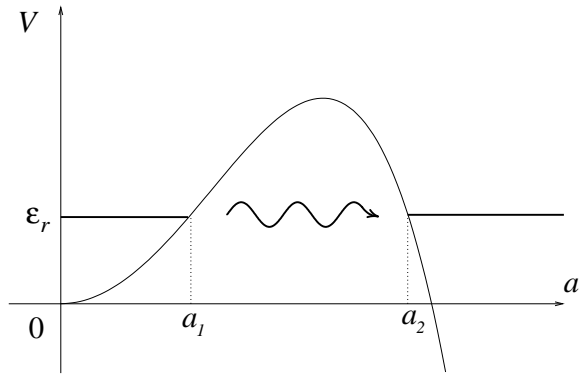}\end{center}

\caption{``Tunneling from something'': creation of an expanding universe
by tunneling.\label{cap:fig1a}}
\end{figure}

It is clear that $\exp (-S/\hbar )$ is a standard WKB
wave function for a particle in a potential $V(a)$ with
energy $\varepsilon _r$. For $H^2\varepsilon _r\ll 1$ the
turning points $a_{1,2}$ are approximated by\begin{eqnarray}
&  & a_1^2\equiv \frac{1-\sqrt{1-4\varepsilon _rH^2}}{2H^2}\approx
\varepsilon _r,\\ &  & a_2^2\equiv \frac{1+\sqrt{1-4\varepsilon
_rH^2}}{2H^2}\approx H^{-2}.
\end{eqnarray}
If $4H^2\varepsilon _r<1$ then $a_1\neq a_2$ are real and
there exist two classically allowed regions and a classically forbidden
region (see Fig.~\ref{cap:fig1a}). The physical picture of the resulting
cosmology is the following. A small closed universe filled with radiation
of energy density $\varepsilon _ra^{-4}$ expands until the maximum
scale factor $a_1$ and then recollapses to $a=0$. An expanding
universe is created with a scale factor $a_2$ by tunneling through
the potential barrier. If the WKB approximation is valid, then the
scale factor $a$ is a semiclassical variable and any monotonic function
of $a$ can be used as a time coordinate in the two classically allowed
Lorentzian regions. The turning point $a=a_2$ corresponds to the
beginning of time in the nucleated expanding universe; the Euclidean
region $a_1<a<a_2$ does not correspond to a classical universe.
If $H=0$, the second turning point is absent (formally $a_2\rightarrow
+\infty $) and there is only one Lorentzian region $0<a<a_1$.

We may rewrite Eqs.~(\ref{eq:S0eq1})-(\ref{eq:Sneq1}) as \begin{eqnarray}
S(a) & = & \pm \int ^a\sqrt{V(a)-\varepsilon _r}\, da,\label{eq:S0sol0}\\
\pm \sqrt{V(a)-\varepsilon _r}S_n' & = & S_n^2-\omega _n^2.\label{eq:Sneq0}
\end{eqnarray}
In the two Lorentzian regions [$0<a<a_1$ and $a>a_2$] the
square root in Eq.~(\ref{eq:S0sol0}) assumes imaginary values (with
a positive imaginary part) and the upper sign denotes a wave $\exp[-S(a)]$
traveling to the right, with
\begin{equation}
S(a)=i\int ^a\sqrt{\varepsilon _r-V(a)}\,
da.\label{eq:Sa-Lorentz}\end{equation}
In the Euclidean region
[$a_1<a<a_2$] the square root in Eq.~(\ref{eq:S0sol0}) is real and the
upper sign denotes an exponentially decaying mode $\exp[-S(a)]$ with
\begin{equation}
S(a)=\int ^a\sqrt{V(a)-\varepsilon _r}\, da.\label{eq:Sa-Euclid}\end{equation}

We shall denote by $S_n(a)$ and $S_n^-(a)$ the
solutions of Eq.~(\ref{eq:Sneq0}) with the upper and the lower signs
respectively, omitting the $^+$ subscript in $S_n^+(a)$
for brevity. If $S(a)$ is given by
Eqs.~(\ref{eq:Sa-Lorentz})-(\ref{eq:Sa-Euclid}), then a linear
combination\begin{eqnarray} \Psi \left(a,\left\{ \chi _n\right\} \right) &
= & C_+\exp \left[-\frac{S(a)}{\hbar }-\sum _nS_n(a)\frac{\chi _n^2}{2\hbar
}\right]\nonumber \\ & + & C_-\exp \left[\frac{S(a)}{\hbar }-\sum
_nS_n^-(a)\frac{\chi _n^2}{2\hbar }\right]\label{eq:psi-sol}
\end{eqnarray}
is an approximate solution of the Wheeler-DeWitt equation. The coefficients
$C_{\pm }$ and the solutions $S_n$, $S_n^-$ must be chosen
separately in each of the three physical domains (two Lorentzian regions
and one Euclidean region). To match these solutions across the turning
points, we need to use the matching conditions of Ref.~\cite{VV88}.
These conditions state, in particular, that the value of $S_n$
must be continuous across the turning points $a_{1,2}$ and
moreover\begin{equation} \lim _{a\rightarrow a_1-0}S_n(a)=\lim
_{a\rightarrow a_1-0}S_n^-(a)=\lim _{a\rightarrow
a_1+0}S_n(a).\label{eq:match-a1}\end{equation} A similar matching condition
must be satisfied at the second turning point $a_2$,\begin{equation}
\lim _{a\rightarrow a_2-0}S_n(a)=\lim _{a\rightarrow a_2-0}S_n^-(a)=\lim
_{a\rightarrow a_2+0}S_n(a).\label{eq:match-a2}\end{equation} Due to the
tunneling boundary condition at $a\rightarrow +\infty $ (no wave traveling
to the left) the coefficient $C_-$ for the region $a>a_2$ must vanish and
we do not need the branch $S_n^-(a)$ in that region. In addition to
Eqs.~(\ref{eq:match-a1})-(\ref{eq:match-a2}) we require that the regularity
condition [Eq.~(\ref{eq:Sn-pos})] holds for $S_n(a)$, $S_n^-(a)$. We
shall refer to such solutions of Eq.~(\ref{eq:Sneq0}) as ``regular''.

Note that the function $S_n^-(a)$ does not have to
be continuous at $a=a_1$. This is because the branch with the lower
signs [the second term in Eq.~(\ref{eq:psi-sol})] is exponentially
small at $a=a_1$ compared with the other branch and cannot be kept
in the WKB approximation.

\subsection{Construction of regular Gaussian solutions}

Our task here is to obtain an explicit solution of\begin{equation}
\sqrt{V(a)-\varepsilon _r}\frac{dS_n}{da}=S_n^2-\omega
_n^2,\label{eq:SnNew}\end{equation} subject to the regularity condition
[Eq.~(\ref{eq:Sn-pos})],\begin{equation} 0<Re\, S_n(a)<\infty \textrm{
for all }a.\end{equation} Here $S_n(a)$ is in general a complex function.
Because of matching conditions, $S_n(a)$ must be continuous
across the turning points $a=a_{1,2}$. Equation~(\ref{eq:SnNew})
is identical to Eq.~(\ref{eq:Sneq0}); the equation for $S_n^-(a)$
can be obtained by changing the sign of the derivative $d/da$ in
Eq.~(\ref{eq:SnNew}).

Following Refs.~\cite{LR79,HH85,Banks},
we introduce the semiclassical time variables: $t$ in the classically
allowed regions and $\tau $ in the under-barrier region, \begin{eqnarray}
\frac{da}{dt} & \equiv  & \sqrt{\varepsilon _r-V(a)},\quad a<a_1\:
\textrm{or}\: a>a_2,\label{eq:t-def}\\ \frac{da}{d\tau } & \equiv  &
\sqrt{V(a)-\varepsilon _r},\quad a_1<a<a_2.\label{eq:tau-def}
\end{eqnarray}
The variable $t$ is the conformal time in a Lorentzian universe,
while $\tau $ is the Euclidean conformal time. Below we shall be using
the variables $a$, $t$ and $\tau $ interchangeably in the appropriate
ranges of $a$. Then Eq.~(\ref{eq:SnNew}) becomes\begin{eqnarray}
i\frac{dS_n}{dt} & = & S_n^2-\omega _n^2,\quad a<a_1\: \textrm{or}\:
a>a_2,\label{eq:Sn-t1}\\ \frac{dS_n}{d\tau } & = & S_n^2-\omega _n^2,\quad
a_1<a<a_2.\label{eq:Sn-tau1}
\end{eqnarray}
The equations for the other branch $S_n^-$ of Eq.~(\ref{eq:Sneq0})
are obtained by reversing the signs of the time derivatives. The branch
$S_n$ corresponds to the wave traveling to the right in the classically
allowed regions and to the decaying underbarrier branch in the Euclidean
region.

To analyze the existence of solutions and to obtain approximations,
it is convenient to transform the function $S_n(a)$
and introduce an auxiliary function\begin{equation}
\zeta _n(a)\equiv \frac{S_n-\omega _n}{S_n+\omega
_n}.\label{eq:xi-def}\end{equation} We imply that the pair of functions
$S_n^{\pm }(a)$ is transformed into a pair $\zeta _n^{\pm }(a)$ but
will often omit the superscripts $^{\pm }$ for brevity. The simple
transformation of Eq.~(\ref{eq:xi-def}) allows to use the functions
$S_n^{\pm }$ and $\zeta _n^{\pm }$ interchangeably. The function
$\zeta _n(t)$ can be interpreted as the ``instantaneous
squeezing parameter'' \cite{JVW1} describing a state of the mode
$\chi _n$ that is related to the vacuum of the instantaneous
diagonalization picture by a time-dependent Bogolyubov transformation.
However, for our purposes it is enough to regard Eq.~(\ref{eq:xi-def}) as a
formal transformation that simplifies calculations.

The regularity condition [Eq.~(\ref{eq:Sn-pos})] is now equivalent
to \begin{equation}
\left|\zeta _n(a)\right|<1.\label{eq:zeta-reg}\end{equation}
This condition is violated by $\left|\zeta _n\right|=1$ when $Re\,
S_n(a)=0$ or by $\zeta _n=1$ when $\left|S_n(a)\right|\rightarrow \infty $.
From Eq.~(\ref{eq:xi-def}) and the matching conditions for $S_n(a)$
it follows that the (complex) function $\zeta _n(a)$
must be continuous across the turning points $a_{1,2}$. From
Eqs.~(\ref{eq:Sn-t1}), (\ref{eq:Sn-tau1}) we obtain the equations for
$\zeta _n(a)$ in each region in terms of the conformal time variables,
\begin{eqnarray} \frac{d}{dt}\zeta _n & = & 2i\omega _n\zeta
_n-\frac{1}{2\omega _n}\frac{d\omega _n}{dt}(1-\zeta
_n^2),\label{eq:zeta-equ-t}\\ \frac{d}{d\tau }\zeta _n & = & 2\omega
_n\zeta _n-\frac{1}{2\omega _n}\frac{d\omega _n}{d\tau }(1-\zeta
_n^2).\label{eq:zeta-equ-tau}
\end{eqnarray}
Our goal now is to obtain a solution for $\zeta _n(a)$
satisfying the regularity condition.

In Appendix~\ref{sec:app-Construction} we derive some technical
results concerning regular solutions of
Eqs.~(\ref{eq:zeta-equ-t})-(\ref{eq:zeta-equ-tau}). In
Appendix~\ref{sub:The-regularity-condition} it is shown that a solution
$\zeta _n(a)$ is regular if $\left|\zeta _n(a_2)\right|<1$, while the
solution $\zeta _n^-(a)$ only needs to be regular at $a=a_1$ to be regular
everywhere. From this it follows that regular solutions of
Eq.~(\ref{eq:Sneq0}) always exist. For instance, one could numerically
integrate Eq.~(\ref{eq:Sneq0}) with the upper sign, starting at $a=a_2$
with any value $S_n(a_2)$ such that $0<Re\, S_n(a_2)<+\infty $, and obtain
a regular solution at $a>a_2$ and at $a<a_2$.

To build a perturbative expansion,
Eqs.~(\ref{eq:zeta-equ-t})-(\ref{eq:zeta-equ-tau}) can be rewritten as
integral equations. For instance, the function $\zeta _n(\tau )$ in the
Euclidean region, with an arbitrary boundary value $\zeta _n(\tau _2)$,
satisfies 
\begin{eqnarray} &&\zeta _n(\tau )  =  \zeta _n(\tau _2)\exp
\left[-2\int _{\tau }^{\tau _2}\omega _nd\tau \right]\nonumber \\ 
& &
+\int _{\tau }^{\tau _2}\frac{\dot{\omega }_n}{2\omega _n}\left[1-\zeta
_n^2(\tau ')\right]\exp \left[-2\int _{\tau }^{\tau '}\omega _nd\tau
\right]d\tau '.
\end{eqnarray}
This equation can be solved iteratively starting with $\zeta _n(\tau
)\equiv 0$. A similar calculation is performed for $\zeta _n(t)$
in the Lorentzian regions. In Appendix~\ref{sub:An-approximate-solution}
we prove that the sequence of iterations always converges.

An important case is an adiabatic (slow) expansion of the universe. The adiabaticity condition is
\begin{equation}
\left | \frac{1}{\omega _n^2}\frac{d\omega
_n}{dt}\right |= \frac{m^2a\sqrt{\left |\varepsilon
_r-V(a)\right |}}{(n^2+m^2a^2)^{3/2}} \ll 1.
\label{eq:adiab}\end{equation}
Analyzing Eq.~(\ref{eq:adiab}) with $V(a)$ given by Eq.~(\ref{eq:Va}), we
find that it holds if $n\gg m/H$ or if $m\gg H$. The only case when the
ratio of Eq.~(\ref{eq:adiab}) is of order $1$ is when $n\sim 1$ and $m\sim H$.

In Appendix~\ref{sub:An-approximate-solution} we also
show under assumption of adiabaticity that there exist solutions
for which
$\left|\zeta _n(a)\right|$ is always small, of order
$\left|\dot{\omega }_n/\omega _n^2\right|\ll 1$, and obtain
the following approximations to general solutions [see also
Eq.~(\ref{eq:zeta-minus-tau})],\begin{eqnarray} \zeta _n(\tau ) & = &
\zeta _n(\tau _2)\exp \left[-2\int _{\tau }^{\tau _2}\omega _nd\tau
\right]\nonumber \\ & + & \int _{\tau }^{\tau _2}\exp \left[-2\int _{\tau
}^{\tau '}\omega _nd\tau \right]\frac{\dot{\omega }_n}{2\omega _n}(\tau
')d\tau ',\label{eq:zeta-plus-0}
\end{eqnarray}
\begin{eqnarray}
\zeta _n^-(\tau ) & = & \zeta _n^-(\tau _1)\exp \left[-2\int _{\tau
_1}^{\tau }\omega _nd\tau \right]\nonumber \\ & - & \int _{\tau _1}^{\tau
}\exp \left[-2\int _{\tau '}^{\tau }\omega _nd\tau \right]\frac{\dot{\omega
}_n}{2\omega _n}(\tau ')d\tau '.\label{eq:zeta-minus-0}
\end{eqnarray}
Here the given boundary values $\zeta _n(\tau _2)$ and $\zeta
_n^-(\tau _1)$ must be
sufficiently small, but are otherwise arbitrary. Similar expressions
hold for the functions $\zeta _n(t)$ and $\zeta _n^-(t)$ in Lorentzian
regions.

Using the above approximations for $\zeta _n$ and $\zeta _n^-$,
we find $S_n^{\pm }$ and find the Gaussian wave function of
Eq.~(\ref{eq:psi-sol}). Calculations verifying the validity of the Gaussian
approximation in the neighborhood of $\chi _n=0$ and the validity of the
WKB approximation for a regular Gaussian solution are given in
Appendix~\ref{sec:app-Applicability}. We conclude that the regular Gaussian
wave function is a valid approximation to an exact solution of the
Wheeler-DeWitt equation.

\subsection{Non-uniqueness of regular wave functions}

\label{sub:Non-uniqueness}The solutions $\zeta _n$, $\zeta _n^-$
of Eqs.~(\ref{eq:zeta-plus-0})-(\ref{eq:zeta-minus-0}) depend
on the arbitrary boundary values $\zeta _n(\tau _2)$
and $\zeta _n^-(\tau _1)$ respectively. The choice
of these boundary values is \emph{a priori} constrained only by regularity
and by matching conditions. The matching condition $\zeta _n(\tau _2)=\zeta
_n^-(\tau _2)$ leaves one free parameter in the resulting wave function.

However, this remaining freedom of choosing a regular solution of
Eq.~(\ref{eq:Sneq0}) does not lead to an appreciable variation in
the resulting functions $S_n^{\pm }(a)$ in the Lorentzian
regions. The properties of Eq.~(\ref{eq:Sneq0}) in the Euclidean
region $a_1<a<a_2$ are such that, as long as the regularity condition
holds, $S_n^-(a_2)$ is almost insensitive to the
initial value $S_n^-(a_1)$, and in turn, the value
$S_n(a_1)$ is almost insensitive to $S_n(a_2)$.
More precisely, the initial value $S_n^-(a_1)$ may
vary throughout a wide range but the final value $S_n^-(a_2)$
is constrained to be within an exponentially small interval.

This can be seen directly from
Eqs.~(\ref{eq:zeta-plus-0})-(\ref{eq:zeta-minus-0}). For instance, the
influence of the boundary condition $\zeta _n(\tau _2)$ on the value $\zeta
_n(\tau _1)$ is suppressed by an exponentially small factor
\begin{equation}
\exp \left[-2\int _{\tau _1}^{\tau _2}\omega _n(\tau )d\tau \right]\equiv
\exp \left[-2\theta _b\right].\label{eq:thetab1}\end{equation}
Here $\theta _b$ is in the
notation of Appendix~\ref{sub:An-approximate-solution} [see
Eq.~(\ref{eq:theta-b-def})]. The quantity $\theta _b$ is never small
since $a\sim H^{-1}$, $V(a)\sim H^{-2}$ and\begin{equation}
\theta _b= \int _{a_1}^{a_2}\frac{\omega
_n(a)da}{\sqrt{V(a)-\varepsilon
_r}}\sim \sqrt{n^2+\frac{m^2}{H^2}}.\label{eq:thetab2}\end{equation}
We find that $\theta _b \gg 1$ when $n\gg 1$ or $m\gg H$.

Since $S_n(a_2)=S_n^-(a_2)$, the
value $S_n(a_1)$ is also constrained to a narrow interval.
Small changes of boundary values
of the functions $S_n(a)$, $S_n^-(a)$ at the turning points $a_{1,2}$
lead to small changes of their values
in the Lorentzian regions. Therefore, different choices
of the free parameter $S_n^-(a_1)$ yield only exponentially
small changes in the resulting solutions within the Lorentzian universes.
The width of the possible range of solutions within a Lorentzian region
is \begin{equation}
\left|\frac{\Delta S_n(t)}{S_n(t)}\right|\sim \exp (-2\theta
_b).\label{eq:thetab3}\end{equation}

A physically interesting simple case is that of a single closed universe,
$H=0$. In that case there is only one Lorentzian region $0<a<a_1$
and the wave function in the Euclidean region has only one branch,
$S_n(\tau )$. In the limit $H\rightarrow 0$, we find
$\theta _b\rightarrow +\infty $. Therefore the influence of the
boundary condition $S_n(a_2)$ at $a_2\rightarrow +\infty $
is completely suppressed and the value $S_n(a_1)$
is uniquely determined by regularity. Similarly, the regular wave
function is unique in the limiting case $\varepsilon _r\rightarrow 0$,
also because $\theta _b\rightarrow +\infty $ [the integral in
Eq.~(\ref{eq:thetab2}) diverges near $a=0$].
[This conclusion does not depend on the adiabaticity condition
implicit in Eq.~(\ref{eq:thetab3});
in Appendix~\ref{sec:unique-infinite} we prove the uniqueness
of the regular solution under much weaker assumptions.]
However, in
the general case ($H\neq 0$ and $\varepsilon _r\neq 0$) the wave
function is not uniquely fixed by the regularity conditions alone.

\subsection{Prescription for a unique Gaussian vacuum}

\label{sub:Prescription}In the previous section we have found a
one-parametric family of well-behaved Gaussian wave functions that can
serve as vacuum states. We may impose an additional constraint on the wave
function to specify it uniquely.

The extra condition is that the branch $S_n^-(a)$
must be continuous across the first turning point $a_1$. Together
with other matching conditions this gives \begin{equation}
S_n^-(a_1)=S_n(a_1).\label{eq:vacuum-unique}\end{equation}
In Appendix~\ref{sub:Uniqueness} we show that this condition is
always satisfied by a unique pair of regular solutions $S_n(a)$,
$S_n^-(a)$. This proves the existence and uniqueness
of the selected state. We shall call this state a ``Gaussian vacuum''.

We view the additional constraint of Eq.~(\ref{eq:vacuum-unique})
as a prescription for defining a unique regular vacuum state. The
wave function resulting from this prescription has the advantage that
it agrees with the vacuum obtained in Ref.~\cite{JVW1} for the massless
scalar field. All other regular solutions obtained in
Sec.~\ref{sub:Non-uniqueness} differ from the Gaussian vacuum by an
exponentially small amount (in Lorentzian regions).

Note that the Gaussian vacuum prescription is not local, since it
requires knowledge of the behavior of the of the frequency
$\omega_n(a)$ and of the potential $V(a)$ everywhere under the barrier.

\section{Interpretation of the Gaussian vacuum and particle production}

\label{sec:Interpretation}A semiclassical Gaussian wave function
is interpreted by making a transition from the Schrödinger picture
of minisuperspace to a QFT in curved spacetime \cite{LR79,HH85,Banks,Wada86}.
Naturally, this can be done only in a Lorentzian region where a
semiclassical spacetime exists. One introduces the conformal time $t$ via
Eq.~(\ref{eq:t-def}). The scalar field $\chi $ is canonically quantized
using the standard creation and annihilation operators ${\mathbf a}_n$,
${\mathbf a}_n^{\dag }$ for the modes $\chi _n$ [here the extra indices
$l$, $p$ of the modes are again suppressed for brevity,
cf.~Eq.~(\ref{eq:field-chi})]. The amplitude $\chi _n$ of the $n$-th
mode is promoted to an operator $\hat{\chi }_n$ and decomposed into
creation and annihilation operators,\begin{equation}
\hat{\chi }_n(t)={\mathbf a}_n\nu _n(t)+{\mathbf a}_n^{\dag }\nu
_n^{*}(t).\end{equation}

A vacuum state is defined by a particular choice of the mode functions
$\nu _n(t)$. If the Gaussian wave function is given
by Eq.~(\ref{eq:Gauss}) with a certain set of $S_n(a)$
for all $n$, then the corresponding mode functions $\nu _n$ are
found from \begin{equation}
S_n(t)=\frac{i}{\nu _n}\frac{d\nu _n}{dt},\quad 0<a<a_1\, \textrm{or}\,
a>a_2.\label{eq:nu-def-t}\end{equation} The interpretation of the Gaussian
wave function [Eq.~(\ref{eq:Gauss})] in the Lorentzian regions is that
the quantized scalar field is in the vacuum state defined by the above mode
functions $\nu _n(t)$.

The mode functions $\nu _n$ are determined by Eq.~(\ref{eq:nu-def-t})
up to an arbitrary constant factor. 
It follows from Eq.~(\ref{eq:Sn-t1}) that $\nu _n(t)$ satisfy
\begin{equation}
\frac{d^2\nu _n}{dt^2}+\omega _n^2\nu
_n=0.\label{eq:nu-equ-t}\end{equation} This is the usual equation for the
mode functions of the $n$-th mode of a (rescaled) conformally coupled
scalar field. The mode functions may be normalized by the Wronskian
condition\begin{equation} \dot{\nu }_n^{*}\nu _n-\dot{\nu }_n\nu
_n^{*}=i.\label{eq:Wronskian}\end{equation}

\subsection{The Gaussian state and the adiabatic vacuum}

In the companion paper \cite{JVW1} we have shown that a Gaussian
wave function satisfying the regularity condition can be interpreted
as a single semiclassical spacetime with a quantized field in a certain
vacuum state. In Sec.~\ref{sub:Prescription} we have defined a unique
vacuum state (the ``Gaussian vacuum'') that satisfies the regularity
condition. Now we need to compare this Gaussian vacuum with a physically
motivated vacuum in our Lorentzian universes.

Neither the spacetime of the recollapsing universe nor that of the
nucleated universe possess a well-defined asymptotically static
regime.  This does not permit one to define unique
``in'' or ``out''
vacuum states of the QFT in these spacetimes. However, one can define an
approximate ``adiabatic vacuum'' of the QFT in a nonstationary FRW
spacetime \cite{PF74} if the adiabatic approximation is valid for the
mode functions of the field.  We now show that the adiabatic vacuum in
fact coincides with the Gaussian vacuum state we defined (up to a
small correction). The difference between the two vacua is in any case
smaller than the uncertainty inherent in the definition of the
adiabatic vacuum.

We assume that the adiabaticity condition holds; its precise formulation
is \begin{equation}
\frac{1}{\omega ^2}\left|\frac{d\omega _n}{dt}\right|\ll 1\end{equation}
[see also Eq.~(\ref{eq:f-adiab})]. In that case one can use
at least a few terms of the WKB expansion for Eq.~(\ref{eq:nu-equ-t}).
To obtain the WKB expansion, one can introduce a formal parameter
$\lambda $, define $\omega _n(t)\equiv \omega _n(\lambda t)$
and use the ansatz\begin{equation}
\nu _n(t)=\frac{1}{\sqrt{W(t)}}\exp \left[-i\int ^t
W(t)dt\right].\label{eq:nu-WKB-0}\end{equation} The WKB function $W(t)$ is
found as an asymptotic series in $\lambda ^2$. The first few terms
are\begin{equation} W=\omega _n-\lambda ^2\left(\frac{\ddot{\omega
}_n}{4\omega _n^2}-\frac{3\dot{\omega }_n^2}{8\omega
_n^3}\right)+...\label{eq:WKB-series}\end{equation} Each derivative of
$\omega _n$ adds a power of $\lambda $; at the end we put $\lambda =1$. The
true small parameter of the expansion is the slowness of change of $\omega
_n(t)$, formalized through $\left|\dot{\omega }_n\right|\ll \omega _n^2$
and analogous conditions on higher time derivatives.

The definition of an adiabatic vacuum depends on the chosen adiabatic
order $k$ and on an arbitrary fiducial time $t_0$. Let the function
$W^{(k)}(t)$ give the WKB series [Eq.~(\ref{eq:WKB-series})]
truncated up to terms $O\left(\lambda ^{2k}\right)$. The mode function
$\nu _n(t)$ describing the adiabatic vacuum are required to coincide
with the WKB solution of order $k$ at $t=t_0$. The condition for
this is\begin{equation}
\left.\frac{\dot{\nu }_n}{\nu _n}+iW^{(k)}+\frac{1}{2}\frac{d}{dt}\ln
W^{(k)}\right|_{t=t_0}=0.\label{eq:adiabatic-vacuum}\end{equation}
In an adiabatically changing background, the adiabatic vacuum of order $k$
gives an apparent particle creation rate of order $k+1$ in the adiabatic
parameter. For a given metric the WKB solution is usable only until a
certain finite order $k_{\max }$ after which the WKB series starts to
diverge.

The condition of Eq.~(\ref{eq:adiabatic-vacuum}) fixes the value
$S_n(t_0)$. [For comparison,
the instantaneous diagonalization approach sets $S_n(t_0)=\omega
_n(t_0)$.] In Appendix~\ref{sub:Asymptotic-adiabatic} we show that the
asymptotic series obtained for $S_n(t)$ in the WKB approximation
is the same (to all orders) as the asymptotic series for $S_n(t)$
obtained from the Gaussian vacuum. Therefore, the Gaussian vacuum
coincides with the adiabatic vacuum within the accuracy of its definition.

This coincidence is not specific to the Gaussian vacuum. Any other
regular Gaussian wave function as found in Sec.~\ref{sub:Non-uniqueness}
will give a value $S_n(t_0)$ different by exponentially
small terms of order $\exp (-2\theta _b)$. The WKB series
contains terms of order $\left[\dot{\omega }_n/\omega _n^2\right]^k$
and is insensitive to exponentially small contributions. In other
words, the definition of an adiabatic vacuum of order $k$ at time
$t_0$ contains inherent uncertainties of much larger magnitude.
We conclude that the regularity of the wave function specifies a quantum
state of the universe that is indistinguishable from an adiabatic
vacuum of any applicable order.

\subsection{Quantum state of the nucleated universe}

Previous work suggests that in the case of tunneling from nothing
($\varepsilon _r=0$) the scalar field in the nucleated universe
should be in the Bunch-Davies (BD) vacuum state. Now we can consider
the case of tunneling from something ($\varepsilon _r\neq 0$) and
compare the Gaussian vacuum state in the asymptotic region $a\rightarrow
+\infty $ with the BD vacuum.

In the Schrödinger picture of minisuperspace, the mode functions $\nu
_n(t)$ of a vacuum are determined by the solutions $S_n(t)$
[Eq.~(\ref{eq:nu-def-t})]. Suppose that another vacuum state
is determined by another set of solutions $\tilde{S}_n(t)$
and the corresponding mode functions $\tilde{\nu }_n(t)$.
The two vacua are related by a Bogolyubov transformation,\begin{equation}
\tilde{\nu }_n(t)=\alpha \nu _n(t)+\beta \nu
_n^{*}(t).\label{eq:x-Bogolyubov}\end{equation} We can express the
Bogolyubov coefficients $\alpha _n$, $\beta _n$ directly through the
functions $S_n(t)$ and $\tilde{S}_n(t)$, as follows. If both sets of mode
functions are normalized [Eq.~(\ref{eq:Wronskian})], then $\alpha _n$
and $\beta _n$ satisfy $\left|\alpha _n\right|^2-\left|\beta
_n\right|^2=1$. We can select any value of $t$ and express $\tilde{\nu
}_n(t)$ and $\dot{\tilde{\nu }}_n(t)$ according to
Eq.~(\ref{eq:x-Bogolyubov}). The solution is\begin{eqnarray}
\alpha _n & = & \dot{\nu }_n^{*}\tilde{\nu }_n-\nu _n^{*}\dot{\tilde{\nu
}}_n,\\ \beta _n & = & -\dot{\nu }_n\tilde{\nu }_n+\nu _n\dot{\tilde{\nu
}}_n.
\end{eqnarray}
Then the physically measurable quantity $\left|\beta _n\right|^2$
that gives the mean occupation number in the mode $\chi_n$
is found (again, independently of a fixed value of $t$) as\begin{equation}
\left|\beta _n\right|^2 = \frac{\left| \tilde{S}_n(t)-S_n(t)\right|^2}{4Re\ \tilde{S}_n(t) Re\ S_n(t)}.\label{eq:S-Bogolyubov}\end{equation}

In the case $\varepsilon _r=0$ the regular solutions $S_n^{\pm }(a)$
of Eq.~(\ref{eq:Sneq0}) are unique and are known to correspond to
the de Sitter-invariant BD vacuum \cite{VV88,GV97}. The difference
between the cases $\varepsilon _r=0$ and $\varepsilon _r\neq 0$
can be seen from Eqs.~(\ref{eq:xi-def}), (\ref{eq:zeta-plus-0}):
the neighborhood of the first turning point $\tau =\tau _1$ contributes
an exponentially small amount to $S_n^-(\tau _2)$.
Therefore the regular solution in the $\varepsilon _r\neq 0$ case
[denoted for now $\tilde{S}_n(a)$] may differ from
the BD solution $S_n(a)$ only by an exponentially small
contribution of order $\exp (-2\theta _b)$. From
Eq.~(\ref{eq:S-Bogolyubov}) we find that the Bogolyubov transformation
relating the two vacua is exponentially close to the identity
transformation ($\alpha _n=1$, $\beta _n=0$).

We conclude that the Gaussian vacuum state derived from the regularity
of the wave function contains an exponentially small number of particles
from the point of view of the BD vacuum.

\section{Discussion}

\label{sec:Discussion}We have shown that a minisuperspace model of
quantum cosmology with a conformally coupled massive scalar field
admits a vacuum state described by a well-behaved Gaussian wave function
(the ``Gaussian vacuum''). Such a vacuum state is provided by
a regular Gaussian solution of the form of Eq.~(\ref{eq:Gauss})
and represents a recollapsing universe and an expanding universe that
is nucleated by quantum tunneling. The obtained wave function describes
a single semiclassical geometry throughout the Lorentzian and Euclidean
regions and does not exhibit any ``branching'' into different
semiclassical geometries. We have checked the consistency of the WKB
approximation, of the Gaussian approximation, and of neglecting the
backreaction of the scalar field perturbations. We have also shown
that the Gaussian vacuum describes both the recollapsing and the expanding
universe in quantum states that are indistinguishable from an adiabatic
vacuum of any meaningful order. 

Unlike the definition of an adiabatic vacuum, our prescription for
the Gaussian vacuum is not local. 
In other words, we cannot specify our preferred quantum state at some value
of $a <a_1$ without a full knowledge of the shape of the barrier $V(a)$
and of the function $\omega _n(a)$ under the barrier.
Nevertheless we believe that our choice of the vacuum state is adequate,
for the following reasons. (i) Our prescription is in the spirit of the
definition of an adiabatic vacuum which was designed to minimize the
apparent particle production. (ii) The Gaussian vacuum coincides with any
adiabatic vacuum within the inherent uncertainty of the latter. (iii) In
the massless case, where a well-motivated independent definition of
the vacuum is available, our prescription selects the correct vacuum
state. (iv) Finally, the Gaussian vacuum represents a fixed semiclassical
background geometry.
Any excited states built from the Gaussian vacuum using a finite number
of creation operators are also well behaved
and represent single geometries.
If we wished to consider tunneling
from a different state of the recollapsing universe, e.g., from a
squeezed state, it would be more illuminating to represent that state
as an infinite superposition of excited states built over the Gaussian
vacuum, with each branch representing a single semiclassical
geometry. The physical interpretation of the QFT in resulting spacetimes
would be unambiguous. It would be interesting if the Gaussian vacuum
(or some state exponentially close to it) could be specified by a set
of local conditions, but at present we are unable to provide such a
specification.

Our construction of the Gaussian vacuum is general and not specific
to the particular physical system we considered. The recent work of
Bouhmadi-López \emph{et~al.}~\cite{BGG02} discusses the quantum
cosmology of a FRW universe filled with massless radiation field and
a conformally coupled massive scalar field, in the presence of a positive
or a negative cosmological constant.
They have used the regularity
conditions to constrain the Gaussian wave functions and obtained a
family of regular Gaussian solutions. This family corresponds to
our family of regular solutions $S_n(a)$, $S_n^-(a)$
parametrized by the boundary value $S_n^-(a_1)$
[Sec.~\ref{sub:Non-uniqueness}]. We have shown in general that all
such regular solutions describe practically the same state of the
Lorentzian universes (up to exponentially small corrections).
Our considerations are more general than those of
Ref.~\cite{BGG02}, where the existence of a family of regular solutions
was demonstrated explicitly in a particular model with a negative
cosmological constant.

Our analysis can also be applied to artificial situations such as
those treated numerically by Rubakov \emph{et~al.}~in
Ref.~\cite{LRR02}. We may consider Eq.~(\ref{eq:SnNew}) with arbitrary
functions $V(a)$, $\omega _n(a)$ as long as the potential $V(a)$ has
the same qualitative behavior as the function of Eq.~(\ref{eq:Va}),
providing a potential barrier. Our statements about the existence and
the behavior of regular solutions hold for any suitably well-behaved
functions $V(a)$ and $\omega _n(a)$. The unique Gaussian vacuum can be
found either by a perturbative expansion or numerically. 

In Ref.~\cite{LRR02} the functions $\omega _n(a)$ and $V(a)$ were
selected so that the resulting Lorentzian universe had an
asymptotic region $a\rightarrow -\infty$, where
$\omega_n(a) \to\omega_n^{(0)}= const$. The asymptotic in-vacuum is then
well-defined and is specified by $S_n(a\to -\infty)= \omega_n^{(0)}$.
However, the Gaussian vacuum for such situations will generally specify
$S_n(a)$ such that\begin{equation}
\lim _{a\rightarrow -\infty }\left[S_n(a)-\omega _n(a)\right]\equiv \Delta
S_n\neq 0.\end{equation} The quantity $\Delta S_n$ can be found using the
methods of Appendix~\ref{sub:An-approximate-solution}. If the adiabaticity
condition is satisfied, all $\Delta S_n$ are exponentially small. 
(Note that the potentials used in Ref.~\cite{LRR02} do not satisfy the
adiabaticity condition.)

Nonzero values of $\Delta S_n$ imply that the in-vacuum is an
infinite superposition of semiclassical wave functions with different
occupation numbers over the Gaussian vacuum.  In fact, it is a
squeezed state with squeezing parameters $\zeta_n\approx \Delta
S_n/2\omega_n^{(0)}$. 
%
Even if states with large particle numbers are exponentially
suppressed to the left of the barrier $a < a_1$,
critical branching may still occur.  The under-barrier
suppression of lower-energy branches is also exponential, and thus one
is faced with a quantitative question of which of the two effects
prevails.  The answer to this question is likely to be model-dependent,
and it is conceivable that a wave function starting in an in-vacuum 
state in the asymptotic region can exhibit critical branching under
the barrier and be dominated by highly excited branches after the tunneling. 
We shall not attempt to address this issue in the present paper.

In the physically realistic case, when there is no asymptotically
static region, the Gaussian vacuum can serve as a
suitable definition of the vacuum state.  The wave function for
``tunneling from nothing'' can then be obtained as a
limit of tunneling from a small recollapsing universe when the energy
of that universe vanishes, $\epsilon_r\to 0$.  In this limit, the
Gaussian vacuum becomes the de Sitter invariant Bunch-Davies vacuum
state. 

\section*{Acknowledgments}

We would like to thank Larry Ford, Jaume Garriga, Alan Guth, Slava
Mukhanov, Matthew Parry, and Takahiro Tanaka for helpful discussions.
We are particularly grateful to Valery Rubakov for stimulating
correspondence.
This work was supported in part by a 2001 Hanyang University Faculty
Research Grant (JH).

\appendix

\section{Properties of a regular vacuum solution}

\label{sec:app-Construction}

\subsection{The regularity condition}

\label{sub:The-regularity-condition}Here we analyze the regularity
condition [Eqs.~(\ref{eq:Sn-pos}) and (\ref{eq:zeta-reg})]
and demonstrate that the regularity condition for all $a$ is equivalent
to imposing the regularity condition only at the turning point. First
we show that if $\left|\zeta _n(t_0)\right|<1$ at
some $t=t_0$ within a Lorentzian region, then $\left|\zeta _n(t)\right|<1$
for all other $t$. Then we demonstrate that the condition $\left|\zeta
_n(\tau )\right|<1$ will hold in the whole Euclidean region $a_1<a<a_2$ if
it holds at the turning point $a_2$. [For the other branch $\zeta
_n^-(\tau )$, a similar argument will show that the regularity condition at
$a=a_1$ is sufficient.] No assumptions of adiabaticity of $\omega _n(a)$
are made.

The function $\zeta _n(a)$ is a differentiable (complex)
function satisfying a first-order differential equation with continuous
coefficients, and a Cauchy problem will have a unique solution. Therefore
each point $\zeta _n^{(0)}$ in the two-dimensional configuration
space (the complex $\zeta _n$ plane) has a unique solution $\zeta _n(a)$
starting from $\zeta _n^{(0)}$ at some $a=a_0$.

First, consider a Lorentzian region (either $0<a<a_1$ or $a>a_2$).
The function $\zeta _n(t)$ satisfies Eq.~(\ref{eq:zeta-equ-t})
and therefore\begin{equation}
Re\left[\zeta _n^{*}\dot{\zeta
}_n\right]=\frac{1}{2}\frac{d}{dt}\left|\zeta
_n\right|^2=-\frac{\dot{\omega }_n}{2\omega _n}\left[1-\left|\zeta
_n\right|^2\right]Re\, \zeta .\end{equation} The quantity $2Re\left[\zeta
_n^{*}\dot{\zeta }_n\right]$ is the ``radial velocity'' at a point $\zeta
_n$. It follows that any trajectory $\zeta _n(t)$ starting on the circle
$\left|\zeta _n\right|=1$ will remain on that circle. From uniqueness,
it follows that no trajectory can cross the unit circle: any solution
$\zeta _n(t)$ that is not entirely on the circle is
either completely inside or completely outside of the circle.

Second, consider a Euclidean region where $\zeta _n(\tau )$
satisfies Eq.~(\ref{eq:zeta-equ-tau}). A similar calculation for
$\left|\zeta _n\right|=1$ gives\begin{equation}
\frac{1}{2}\frac{d}{d\tau }\left|\zeta _n\right|^2=2\omega _n\left|\zeta
_n\right|^2>0\textrm{ at }\left|\zeta _n\right|=1.\end{equation} Therefore
all trajectories $\zeta _n(\tau )$ that cross the unit circle must go
outwards at the crossing point. The condition $\left|\zeta _n(\tau
_2)\right|<1$ guarantees $\left|\zeta _n(\tau )\right|<1$ for all $\tau
<\tau _2$. If we take $\tau _2$ to be the value of $\tau $ at the second
turning point $a=a_2$, then the required statement $\left|\zeta
_n(a)\right|<1$ for $a_1<a<a_2$ follows.

\subsection{An approximate solution}

\label{sub:An-approximate-solution}The main statement of this section
is the following. If the adiabaticity condition \begin{equation}
\left|\frac{1}{2\omega _n^2}\frac{d\omega _n}{dt}\right|\ll 1,\quad
\left|\frac{1}{2\omega _n^2}\frac{d\omega _n}{d\tau }\right|\ll
1,\end{equation} is satisfied and $\omega _(t)$, $\omega_ (\tau )$ are
slowly-changing and non-oscillating functions,
then there exists a regular solution $\zeta
_n$ of Eqs.~(\ref{eq:zeta-equ-t}), (\ref{eq:zeta-equ-tau}) which is always
small, $\left|\zeta _n\right|\ll 1$. The proof is constructive
and uses an integral equation for $\zeta _n$ to build a perturbative
expansion. The method is based on a time-dependent perturbation theory;
a somewhat similar but more cumbersome treatment is in Ref.~\cite{Parker}.
As a by-product we shall find \emph{all} regular solutions $\zeta _n(a)$,
parametrized by the boundary condition at $a=a_2$.

It will be convenient to consider one equation for $\zeta _n$ as
a function of one complex variable instead of two
Eqs.~(\ref{eq:zeta-equ-t})-(\ref{eq:zeta-equ-tau}) using two different time
variables. Start with Eq.~(\ref{eq:SnNew}) and define the new ``time''
variable $\theta $ by\begin{equation} \theta (a)\equiv \int
_{a_2}^a\frac{\omega _n(a)}{\sqrt{\varepsilon
_r-V(a)}}da.\label{eq:theta-def}\end{equation} Here the square root has the
standard branch cut, $Re\, \sqrt{z}>0$, and the contour integration in
Eq.~(\ref{eq:theta-def}) uses $a\rightarrow a+i\delta $ with $\delta >0$ at
the poles $a=a_{1,2}$. The value of $\theta $ is real and positive for
$a>a_2$ and coincides with the phase $\theta (t)=\int _{t_2}^t\omega _ndt$.
In the Euclidean region ($a_1<a<a_2$) we obtain, with the above branch cut,
$\theta (\tau )=i\int _{\tau }^{\tau _2}\omega _nd\tau $. In the Lorentzian
region $0<a<a_1$ we have $\theta (t)=i\theta _b+\int _t^{t_1}\omega _ndt$,
where \begin{equation} \theta _b\equiv \int _{\tau _1}^{\tau _2}\omega
_nd\tau =\int _{a_1}^{a_2}\frac{\omega _n(a)da}{\sqrt{V(a)-\varepsilon
_r}}\label{eq:theta-b-def}\end{equation} is the under-barrier ``Euclidean
phase'' and $t_{1,2}$, $\tau _{1,2}$ are the values corresponding to
$a=a_{1,2}$. For our purposes it is enough to consider only the half-plane
$Im\, \theta \geq 0$.

Denote for convenience\begin{equation}
f(a)\equiv \frac{1}{2\omega _n^2}\frac{d\omega
_n}{dt}=\frac{m^2a\sqrt{\varepsilon
_r-V(a)}}{2(n^2+m^2a^2)^{3/2}}.\label{eq:f-adiab}\end{equation} If the
adiabaticity condition is satisfied, the value of $f$ is everywhere small
and we can find a bound $f_0$ such that $\left|f(a)\right|<f_0\ll 1$.

The equation for $\zeta _n(\theta )$ is\begin{equation}
\frac{d\zeta _n}{d\theta }=2i\zeta _n-(1-\zeta _n^2)f(\theta
).\label{eq:zeta-equ-theta}\end{equation}

From our analysis in Appendix~\ref{sub:The-regularity-condition}
it follows that a solution $\zeta _n(a)$ which is regular,
$\left|\zeta _n(a)\right|<1$ for all $a$, is uniquely
determined by its value at the second turning point $a=a_2$ (i.e.~at
$\theta =0$). Therefore we assume a boundary condition $\zeta
_n\left(\theta =0\right)=b$ with an arbitrary $\left|b\right|<1$. Then the
solution of Eq.~(\ref{eq:zeta-equ-theta}) satisfies an integral
equation,\begin{equation} \zeta _n(\theta )=be^{2i\theta }-\int _0^{\theta
}e^{2i(\theta -\theta ')}\left(1-\zeta _n^2(\theta ')\right)f(\theta
')d\theta '.\label{eq:zeta-theta-integral}\end{equation} The integration in
Eq.~(\ref{eq:zeta-theta-integral}) is understood as contour integration
along e.g.~a straight line connecting $0$ and $\theta $. [The function
$f(a)$ has a branch point at $a=in/m$ and we assume that an appropriate
branch cut is imposed in the complex $\theta $ plane.]

We assume that $\left|f(\theta )\right|<f_0$ for the
relevant values of $\theta $, where $f_0$ is a fixed number. Then
the integral equation for $\zeta _n(\theta )$ can be
solved iteratively, starting with $\zeta _n\equiv 0$. This corresponds
to a perturbation theory expansion in $f_0$ for
Eq.~(\ref{eq:zeta-equ-theta}).

We now show that the iteration of Eq.~(\ref{eq:zeta-theta-integral})
always converges as long as $Im\, \theta >0$. Denote $\zeta _n^{(k)}(\theta
)$ the $k$-th element of the iteration sequence. The initial function
is\begin{equation}
\zeta _n^{(0)}(\theta )=be^{2i\theta }-\int _0^{\theta }e^{2i(\theta
-\theta ')}f(\theta ')d\theta ',\label{eq:zeta-ans1}\end{equation} and the
next approximations are found as\begin{equation} \zeta _n^{(k+1)}(\theta
)=\zeta _n^{(0)}(\theta )+\int _0^{\theta }e^{2i(\theta -\theta
')}\left[\zeta _n^{(k)}(\theta ')\right]^2f(\theta ')d\theta
'.\label{eq:zeta-sequence}\end{equation} The difference between successive
approximations is\begin{eqnarray} & \zeta _n^{(k+1)} & -\zeta
_n^{(k)}=\int _0^{\theta }d\theta '\, f(\theta ')e^{2i(\theta -\theta
')}\nonumber \\ &  & \times \left(\left[\zeta _n^{(k)}(\theta
')\right]^2-\left[\zeta _n^{(k-1)}(\theta
')\right]^2\right).\label{eq:zeta-diff}
\end{eqnarray}
Now we shall estimate the integral in Eq.~(\ref{eq:zeta-diff}) and
show that the LHS tends to $0$ as $k\rightarrow \infty $.

From $0<Im\, \theta '<Im\, \theta $ we get $\left|\exp \left[2i(\theta
'-\theta )\right]\right|\leq 1$. Since $\left|\zeta _n(\theta )\right|<1$
and $\left|f(\theta )\right|\leq f_0$, we can estimate \begin{eqnarray}
&  & \left|\zeta _n^{(k+1)}(\theta )-\zeta _n^{(k)}(\theta
)\right|\nonumber \\ & < & f_0\int _0^{\theta }\left|\left[\zeta
_n^{(k)}(\theta ')\right]^2-\left[\zeta _n^{(k-1)}(\theta
')\right]^2\right|d\theta '\nonumber \\ & < & 2f_0\int _0^{\theta
}\left|\zeta _n^{(k)}(\theta ')-\zeta _n^{(k-1)}(\theta ')\right|d\theta
'.\label{eq:zeta-diff-1}
\end{eqnarray}
Starting from $\left|\zeta _n^{(1)}(\theta )-\zeta _n^{(0)}(\theta
)\right|<f_0\left|\theta \right|$, we can prove by induction that
\begin{equation} \left|\zeta _n^{(k)}(\theta )-\zeta _n^{(k-1)}(\theta
)\right|<\frac{\left(2f_0\left|\theta
\right|\right)^k}{2k!}.\label{eq:zeta-conv}\end{equation} This sequence
clearly tends to $0$ as $k\rightarrow \infty $ at fixed $\theta $.

In terms of real time variables, Eq.~(\ref{eq:zeta-theta-integral})
can be rewritten as e.g.~for the $S_n^-$ branch in the Euclidean
region,\[
\zeta _n^-(\tau )=\zeta _n^-(\tau _1)\exp \left[-2\int _{\tau _1}^{\tau
}\omega _nd\tau \right]\]
\begin{equation}
-\int _{\tau _1}^{\tau }\exp \left[-2\int _{\tau '}^{\tau }\omega _nd\tau
\right]\left(1-\left[\zeta _n^-(\tau ')\right]^2\right)f(\tau ')d\tau
'.\label{eq:zeta-minus-tau}\end{equation}

We have obtained the solution $\zeta _n(\theta )$ of
Eq.~(\ref{eq:zeta-equ-theta}) for the boundary condition $\zeta
_n\left(\theta =0\right)=b$ (with arbitrary $\left|b\right|<1$) as the
limit \begin{equation} \zeta _n(\theta )=\lim _{k\rightarrow \infty }\zeta
_n^{(k)}(\theta )\end{equation} of a sequence $\zeta _n^{(k)}$ defined by
Eqs.~(\ref{eq:zeta-ans1})-(\ref{eq:zeta-sequence}).

Now we can show that if a boundary value is small, $\left|b\right|\ll f_0$,
then $\zeta _n$ is always small and at most of order $f_0$ in the adiabatic case $f_0\ll 1$.
The first iteration [Eq.~(\ref{eq:zeta-ans1})] clearly satisfies
$\left|\zeta _n^{(0)}(\theta )\right|\sim f_0$; each
subsequent iteration will only add terms of higher order in $f_0$.
Therefore, the dominant term of the solution is given by
Eq.~(\ref{eq:zeta-ans1}).
Since
$Im\,\theta\geq0$, it is clear from Eq.~(\ref{eq:zeta-ans1}) that $\left|\zeta_{n}(\theta)\right|$
is of order $\left|f_{0}\right|$ everywhere. We now only need to check that
the oscillating integral in Eq.~(\ref{eq:zeta-ans1}) does not accumulate a
large value when $\theta$ is large. {[}This could happen, for example, in the
case of a parametric resonance when $\omega(t)$ is oscillating.{]}
By assumption $f(\theta)$ is itself a slowly-changing function of
$\theta$ and we can approximate $f(\theta)\approx f_{1}+f_{2}\theta$
where $f_{1}$ and $f_{2}\equiv df/d\theta$ are small constants of order $f_{0}$.
Then the integral over one oscillation between $\theta$ and $\theta+2\pi$
is \begin{equation}
\int_{0}^{2\pi}(f_{1}+f_{2}\theta)e^{-2i\theta}d\theta=i\pi f_{2}.\end{equation}
The value accumulated over many oscillations between $\theta_{1}$ and $\theta_{2}\equiv\theta_{1}+2\pi k$
can be approximated as \begin{equation}
\sum_{j=0}^{k-1}i\pi\frac{df}{d\theta}(\theta_{1}+2\pi j)\approx\frac{i}{2}\int_{\theta_{1}}^{\theta_{2}}\frac{df}{d\theta}d\theta=\frac{i}{2}\left[f(\theta_{2})-f(\theta_{1})\right].\end{equation}
This value is bounded by $\left|f_{0}\right|$. Therefore the integral of Eq.~(\ref{eq:zeta-ans1})
remains small also for large values of $\theta$.

\subsection{An asymptotic series for the adiabatic case}

\label{sub:Asymptotic-adiabatic}We can use
Eq.~(\ref{eq:zeta-theta-integral}) to obtain an asymptotic expansion for
$\zeta _n(\theta )$ if the adiabaticity condition $f_0\ll 1$ holds. We
shall also assume that $b\sim f_0$. The expansion is in the number of
derivatives in $\theta $ as well as in powers of $f$ and is found through
integration by parts, e.g.\begin{equation}
\int _0^{\theta }e^{2i(\theta -\theta ')}f(\theta ')d\theta '\sim
-\left.\sum _{n=0}^{\infty }\frac{e^{2i\theta
}}{(2i)^{n+1}}\frac{d^nf(\theta )}{d\theta ^n}\right|_0^{\theta
}.\label{eq:f-asympt}\end{equation} Similarly to the argument leading to
Eq.~(\ref{eq:zeta-diff-1}), one can show that the iteration sequence of
Eq.~(\ref{eq:zeta-theta-integral}) gives a convergent sequence of
asymptotic series in which the $k$-th iteration changes only terms of order
$k$ and higher. Therefore the asymptotic series for the solution $\zeta
_n(\theta )$ is the limit of this sequence. [Each asymptotic series in
the sequence, like the series in Eq.~(\ref{eq:f-asympt}), may not actually
converge.] In the Lorentzian region $0<a<a_1$ we can omit the
exponentially suppressed terms proportional to $\exp (2i\theta )$. Then
the first terms of the resulting asymptotic series are\begin{equation}
\zeta _n(\theta )\sim
\frac{f}{2i}+\frac{f'}{(2i)^2}+\frac{f''-f^3}{(2i)^3}+\frac{f'''-5f'f^2}{(2i)^4}+...\end{equation} (here the prime denotes $d/d\theta $ and all derivatives of $f$
are evaluated at the same point $\theta $). Converted back to the
time variable $t$, this becomes\begin{eqnarray}
\zeta _n(t) & \sim  & -\frac{i}{4}\frac{\dot{\omega }_n}{\omega
_n^2}-\frac{1}{8}\left(\frac{\ddot{\omega }_n}{\omega
_n^3}-2\frac{\dot{\omega }_n^2}{\omega _n^4}\right)\nonumber \\ &  &
+\frac{i}{16}\left(\frac{\dddot{\omega }_n}{\omega _n^4}-7\frac{\dot{\omega
}_n\ddot{\omega }_n}{\omega _n^5}\right)+...\label{eq:zeta-asympt-ans1}
\end{eqnarray}
This asymptotic series is a power series in the adiabatic parameter
and necessarily misses any exponentially small contributions.

\subsection{Asymptotic series from the WKB approximation}

\label{sub:Asymptotic-series}In the adiabatic case the WKB approximation
may be applied to Eq.~(\ref{eq:nu-equ-t}) to obtain a
solution\begin{equation} \nu _n(t)\propto \frac{1}{\sqrt{W}}\exp
\left[-i\int ^tWdt\right].\label{eq:nu-WKB}\end{equation} The auxiliary
function $W(t)$ is found as an asymptotic WKB series. One may iterate the
equation\begin{equation} W^{(k+1)}=\sqrt{\omega
_n^2-\frac{1}{2}\frac{\ddot{W}^{(k)}}{W^{(k)}}+\frac{3}{4}\left[\frac{\dot{W}^{(k)}}{W^{(k)}}\right]^2},\end{equation} starting from $W^{(0)}\equiv \omega _n(\lambda t)$
and expanding in the formal adiabatic parameter $\lambda $. At the
end one sets $\lambda =1$. Then the solution $S_n(t)$
is obtained from Eqs.~(\ref{eq:nu-def-t}), (\ref{eq:nu-WKB-0})
and transformed into $\zeta _n(t)$ using Eq.~(\ref{eq:xi-def}).
The first terms of the resulting series are\begin{eqnarray}
S_n(t) & \sim  & \omega _n+\frac{\dot{\omega }_n}{2i\omega
}-\left(\frac{\ddot{\omega }_n}{4\omega _n^2}-\frac{3}{8}\frac{\dot{\omega
}_n^2}{\omega _n^3}\right)+...,\label{eq:Sn-asympt-ans2}\\ \zeta _n(t) &
\sim  & -\frac{i}{4}\frac{\dot{\omega }_n}{\omega
_n^2}-\frac{1}{8}\left(\frac{\ddot{\omega }_n}{\omega
_n^3}-2\frac{\dot{\omega }_n^2}{\omega
_n^4}\right)+...\label{eq:zeta-asympt-ans2}
\end{eqnarray}

It is clear that Eq.~(\ref{eq:zeta-asympt-ans2}) should coincide
with the asymptotic series obtained above in
Eq.~(\ref{eq:zeta-asympt-ans1}) using a different approach. An asymptotic
expansion in powers of the adiabatic parameter will necessarily miss any
exponentially small contributions, due to the nature of the power-law
expansion. The remaining asymptotic series is unique for a given function
$\zeta _n(t)$, whether it was obtained from a WKB expansion or from any
other procedure. An advantage of the method of
Appendix~\ref{sub:Asymptotic-adiabatic} is that it can compute
exponentially small terms in the solution.

\subsection{Existence and uniqueness of the Gaussian vacuum}

\label{sub:Uniqueness}Here we prove that for any continuous function
$\omega _n(\tau )$ 
on an interval $\left [\tau_1, \tau_2\right ]$
there exists a unique pair of (complex)
functions $S_n(\tau )$, $S_n^-(\tau )$
that satisfy the equations\begin{eqnarray}
\frac{dS_n}{d\tau } & = & S_n^2-\omega _n^2,\label{eq:Sn-tau-equ}\\
-\frac{dS_n^-}{d\tau } & = & \left[S_n^-\right]^2-\omega
_n^2,\label{eq:Sn-minus-tau-equ}
\end{eqnarray}
the regularity conditions for any $\tau \in \left[\tau _1,\tau
_2\right]$,\begin{equation} 0<Re\, S_n(\tau ),\: Re\, S_n^-(\tau )<+\infty
\end{equation} and the matching conditions\begin{equation}
S_n^-(\tau _1)=S_n(\tau _1),\quad S_n^-(\tau _2)=S_n(\tau
_2).\end{equation}

Consider an auxiliary function $g(s,\tau )$ defined as
the value $g(s,\tau )\equiv S_n(\tau )$ obtained
by solving Eq.~(\ref{eq:Sn-tau-equ}) with the boundary condition $s=S_n(\tau
_2)$. Due to the uniqueness theorem for first-order differential
equations, the function $g(s,\tau )$ is differentiable and provides
a one-to-one map of the complex $s$ plane at any fixed $\tau $.
It is also clear that $g(s,\tau )$ has real values for
real $s$. From our results in Appendix~\ref{sub:The-regularity-condition}
it follows that $0<Re\, g(s,\tau )<+\infty $ for any $s$
such that $0\leq Re\, s<+\infty $ and for $\tau _1\leq \tau <\tau _2$.
The final useful property of $g(s,\tau )$ is $\left|\partial g/\partial
s\right|<1$ for $0<Re\, s<+\infty $. We can prove it as follows. The
function $\partial g(s,\tau )/\partial s$ as a function of $\tau $
satisfies \begin{equation}
\frac{d}{d\tau }\frac{\partial g}{\partial s}=2g(s,\tau )\frac{\partial
g}{\partial s},\quad \frac{\partial g}{\partial s}\left(\tau =\tau
_2\right)=1.\end{equation} This equation determines the function $\partial
g/\partial s$ at fixed $s$ as\begin{equation}
\frac{\partial g}{\partial s}(\tau )=\exp \left[-2\int _{\tau }^{\tau
_2}g(s,\tau )d\tau \right].\end{equation} Since $Re\, g(s,\tau )>0$, we
obtain $\left|\partial g/\partial s\right|<1$ for any $\tau \in \left[\tau
_1,\tau _2\right]$. This means, in particular, that the map $s\rightarrow
g(s,\tau _1)$ decreases the distance between points in the complex $s$
plane.

Similarly, we define the function $g^-$ by solving
Eq.~(\ref{eq:Sn-minus-tau-equ}) with the boundary condition $S_n^-(\tau
_1)=s$ to find $S_n^-(\tau )\equiv g^-(s,\tau )$.
The function $g^-(s,\tau )$ has the same properties
as $g(s,\tau )$.

The problem of finding $S_n$, $S_n^-$ is now equivalent to
solving the algebraic equation $g(g^-(s,\tau _1),\tau _2)=s$
for $s$. The map $s\rightarrow g(g^-(s,\tau _1),\tau _2)$
clearly has the same properties as the functions $g$ and $g^-$.
The existence of a real positive solution $s>0$ follows from the
fact that the functions are continuous, have real values for real
$s$, and satisfy $0<g(s,\tau ),g^-(s,\tau )<+\infty $
for $0\leq s<+\infty $. The solution is unique because if we assume
that $s_1\neq s_2$ are two solutions, then the distance between
$s_1$ and $s_2$ is decreased after applying the map $s\rightarrow
g\left(g^-(s,\tau _1),\tau _2\right)$, which is a contradiction since by
assumption $s_{1,2}$ are stationary points of this map.
[The unique solution can be obtained numerically by iterating the map.]

\subsection{Uniqueness of the regular solution for infinite barriers}

\label{sec:unique-infinite}
Here we consider the equation\begin{equation}
\frac{dS}{d\tau}=S^2-\omega^2(\tau)\label{eq:Sequ-pure}\end{equation}
on the interval $0\leq\tau<+\infty$. We call a solution $S(\tau)$
regular if $0<Re\, S(\tau)<+\infty$ for all $\tau\geq0$.
The main statement is that Eq.~(\ref{eq:Sequ-pure}) has a unique
regular solution $S(\tau)$ if the function $\omega^2(\tau)$
is continuous and bounded from below at large enough $\tau$, namely
$\omega(\tau)\geq f$ at $\tau>\tau_0$, with a suitable
constant $f>0$.

The technical conditions of this statement [continuity and a lower
bound on $\omega(\tau)$] are sufficient but not necessary.
The continuity of $\omega(\tau)$
is used only to ensure that the Cauchy problem for
Eq.~(\ref{eq:Sequ-pure}) has a unique solution that is a continuous
function of the initial condition.

It is enough to consider the case $\omega(\tau)\geq f$
for all $\tau>0$ because a unique regular solution in a domain
$\tau>\tau_0$ is uniquely extended to a regular solution for
$0\leq\tau\leq\tau_0$.

Firstly, we prove that any regular solution $S(\tau)$
must satisfy \begin{equation}
Re\, S(\tau)\geq f\label{eq:S-bound-1}\end{equation}
for all $\tau\geq0$. For this it is enough to show that a regular
solution satisfies $Re\, S(\tau)>f-\varepsilon$ for arbitrary
$\varepsilon>0$. The differential equation for $R(\tau)\equiv Re\, S(\tau)$
is\begin{equation}
\frac{dR}{d\tau}=R^2-\left[Im\, S(\tau)\right]^2-\omega^2(\tau)\leq
R^2-f^2.\end{equation} So the function $R(\tau)$ cannot grow faster than a
solution $R_0(\tau)$ of the equation\begin{equation}
\frac{dR_0}{d\tau}=R_0^2-f^2.\label{eq:R0-equ}\end{equation}
Given a regular solution $S(\tau)$ and a point $\tau_1$,
we can choose $R_0(\tau)$ such that $R_0(\tau_1)\leq Re\, S(\tau_1)$
and then it follows that $R_0(\tau)$ is a lower bound
for $Re\, S(\tau)$ for $0\leq\tau<\tau_1$. Since we
know that $Re\, S(\tau)>0$, we can use the boundary condition
$R_0(\tau_1)=0$ at any $\tau_1>0$ to obtain lower
bounds on $Re\, S(\tau)$. The general solution of Eq.~(\ref{eq:R0-equ})
is\begin{equation}
R_0(\tau)=f\frac{1-Ae^{2f\tau}}{1+Ae^{2f\tau}},\end{equation}
where $A$ is an integration constant. It is easy to see that any
solution $R_0(\tau)$ equal to zero at some large $\tau$
must be exponentially close to $f$ at smaller $\tau$.
More precisely, for any $\varepsilon$
such that $0<\varepsilon<f$ and for any $\tau_1$ the solution
$R_0(\tau)$ will satisfy $R_0(\tau)>f-\varepsilon$
for $0\leq\tau<\tau_1$ if we impose the boundary condition $R_0(\tau_2)=0$
at\begin{equation}
\tau_2=\tau_1+\frac{1}{2f}\ln\frac{2f-\varepsilon}{\varepsilon}.\end{equation}
But the range of $\tau$ is infinite and we can chose $\tau_1$
to be arbitrarily large. Since $R_0(\tau)$ is a lower
bound for $Re\, S(\tau)$, it follows that $Re\, S(\tau)>f-\varepsilon$
for $0\leq\tau<\tau_1$ with any $\tau_1>0$. Therefore, we have
shown that any regular solution of Eq.~(\ref{eq:Sequ-pure}) satisfies
Eq.~(\ref{eq:S-bound-1}) for all $\tau$.

Secondly, we prove that the relative difference between any two regular
solutions $S_1(\tau)$, $S_2(\tau)$ decreases
exponentially with diminishing $\tau$. More precisely, for any two
values $\tau_a$, $\tau_b$ such that $\tau_a<\tau_b$, \begin{equation}
\frac{\left|S_1(\tau_a)-S_2(\tau_a)\right|^2}{\left|S_1(\tau_a)\right|^2+\left|S_2(\tau_a)\right|^2}\leq\frac{\left|S_1(\tau_b)-S_2(\tau_b)\right|^2}{\left|S_1(\tau_b)\right|^2+\left|S_2(\tau_b)\right|^2}e^{-2f(\tau_b-\tau_a)},\label{eq:S-bound-2}\end{equation}
as long as these solutions satisfy $Re\, S_{1,2}(\tau)\geq f$
for $\tau_a\leq\tau\leq\tau_b$. This can be proved directly by
using Eqs.~(\ref{eq:Sequ-pure})-(\ref{eq:S-bound-1}) to
evaluate\begin{eqnarray} &  &
\frac{d}{d\tau}\ln\frac{\left|S_1(\tau)-S_2(\tau)\right|^2}{\left|S_1(\tau)\right|^2+\left|S_2(\tau)\right|^2}\nonumber \\ & = & 2\frac{\left|S_1\right|^2Re\, S_2+\left|S_2\right|^2Re\, S_1+\omega^2Re(S_1+S_2)}{\left|S_1(\tau)\right|^2+\left|S_2(\tau)\right|^2}\nonumber \\
& \geq & 2f.\end{eqnarray}

From Eqs.~(\ref{eq:S-bound-1}) and (\ref{eq:S-bound-2}) we find
that the difference between any two regular solutions at $\tau=0$
must be equal to zero. This follows from
Eq.~(\ref{eq:S-bound-2}) with $\tau_a =0$: the relative difference
by definition cannot exceed $1$, and the RHS can be made arbitrarily small
by choosing large enough $\tau_b$. Therefore the regular solution is unique.

The existence of a regular solution can be proved by construction.
We find a regular solution $S(\tau)$ as the limit of a
sequence of solutions that are regular for a finite part of the interval
$0\leq\tau<+\infty$. Consider a (real) function $S\left(\tau;\tau_0\right)$
defined as the solution of Eq.~(\ref{eq:Sequ-pure}) with the boundary
condition $S\left(\tau_0;\tau_0\right)=0$. The function
$S\left(\tau;\tau_0\right)$ is not a regular solution because it satisfies
$Re\, S(\tau)>0$ only on the interval $0\leq\tau<\tau_0$ but not at larger
$\tau$. We now show that the limit of $S\left(\tau;\tau_0\right)$ as
$\tau_0\rightarrow+\infty$ is a regular solution. The existence of the
pointwise limit (taken separately at each $\tau$) \begin{equation}
S(\tau)\equiv\lim_{\tau_0\rightarrow+\infty}S\left(\tau;\tau_0\right)\label{eq:S-limit}\end{equation}
follows from Eq.~(\ref{eq:S-bound-2}): the relative difference between
functions becomes exponentially small when $\tau_0$ grows. It is
clear that the resulting function $S(\tau)$ will satisfy
Eq.~(\ref{eq:S-bound-1}). It remains to show that the function $S(\tau)$
defined by Eq.~(\ref{eq:S-limit}) is actually a solution of
Eq.~(\ref{eq:Sequ-pure}). This follows from a continuity argument. A
solution of Eq.~(\ref{eq:Sequ-pure}) is a continuous function of an initial
condition. Therefore the limit of Eq.~(\ref{eq:S-limit}) is the same as the
solution of Eq.~(\ref{eq:Sequ-pure}) with the boundary
condition\begin{equation}
S\left(\tau=0\right)=\lim_{\tau_0\rightarrow+\infty}S\left(\tau=0;\tau_0\right).\end{equation} This completes the proof of the existence and uniqueness of the regular
solution of Eq.~(\ref{eq:Sequ-pure}).

\section{Applicability of the approximations}

\label{sec:app-Applicability}In this Appendix we analyze the applicability
of the Gaussian and WKB approximations to the Wheeler-DeWitt equation.
We use the WKB approximation in two places in our calculation. First,
the WKB approximation is used in substituting the Gaussian ansatz
into the Wheeler-DeWitt equation, when all terms of order $O(\hbar )$
are ignored. Second, the WKB approximation is applied to
Eq.~(\ref{eq:nu-equ-t}). We analyze the latter approximation first.

\subsection{Using the WKB approximation for Eq.~(\ref{eq:nu-equ-t})}

\label{sub:WKB}The WKB ansatz of Eq.~(\ref{eq:nu-WKB})
is a good approximation for
Eq.~(\ref{eq:nu-equ-t}) as long as \begin{equation}
\dot{\omega }_n\ll \omega _n^2.\end{equation}
Comparing this with Eq.~(\ref{eq:zeta-asympt-ans1}), we find that
the squeezing parameter $\zeta _n(a)$ is small if and
only if the WKB approximation is valid.

Consider the behavior of $\zeta _n(a)$ in the Euclidean
region where we use the time variable $\tau $ (overdots will denote
derivatives by $\tau $). Taking the leading term of
Eq.~(\ref{eq:zeta-asympt-ans1}),\begin{equation} \zeta _n(a)\approx
\frac{\dot{\omega }_n}{4\omega _n^2}=\frac{m^2a\sqrt{V(a)-\varepsilon
}}{4(n^2+m^2a^2)^{3/2}}\label{eq:zeta-eq1}\end{equation} and estimating
$V(a)-\varepsilon \lesssim a^2$, we obtain\begin{equation} \zeta
_n(a)<\frac{m^2a^2}{4(n^2+m^2a^2)^{3/2}}.\label{eq:zeta-est1}\end{equation}
At fixed $n$, this function reaches a maximum at $ma=n\sqrt{2}$, which
gives a bound\begin{equation} \zeta
_n(a)<\frac{1}{2n\sqrt{27}}.\end{equation} Therefore $\left|\zeta
_n(a)\right|\ll 1$ and the WKB approximation is valid if $n\gg 1$.
Similarly, one can show that the WKB is valid if $ma\gg n$ or if $ma\ll n$.
The only case when the WKB may not apply to Eq.~(\ref{eq:zeta-eq1}) is when
$n\sim 1$ and $ma\sim 1$ at the same time; the maximum value of $\zeta _n$
is at most $\approx 0.1$ in this case.

In the Lorentzian region, Eq.~(\ref{eq:zeta-eq1}) gives at large $a$
\begin{equation}
\left| \zeta_n\right| \sim \frac{H}{4m}.
\end{equation}
The adiabaticity condition is satisfied at large $a$ only if $H\ll m$.
However, we know from Appendix~\ref{sub:The-regularity-condition} that
the solution $\zeta_n(a)$ remains regular in Lorentzian regions even if
the adiabadicity condition does not hold somewhere in those regions.

\subsection{Using the WKB approximation for the WDW equation}

When we substitute the Gaussian ansatz into the Wheeler-DeWitt equation
[Eq.~(\ref{eq:wdw})], we disregard terms of order $O(\hbar )$.
These terms are $\hbar S_0''$, $\hbar S_n''$, and the ``backreaction''
term $\hbar \sum _nS_n$. The WKB approximation is valid if the
disregarded terms are smaller than the typical magnitude of other
terms in the respective equation. For the first two terms, the conditions
are\begin{equation}
\hbar S_0''\ll V(a),\quad \hbar S_n''\ll \omega
_n^2.\label{eq:WKB-a}\end{equation} The WKB approximation may only be valid
away from the turning points. We find \begin{equation}
\left|\frac{S_0''}{V}\right|=\frac{\left|V'\right|}{2V^{3/2}}\approx
H^2\frac{\left|1-2\lambda ^2\right|}{\lambda ^2(1-\lambda
^2)^{3/2}};\end{equation} this is small when $H\ll 1$ away from turning
points. We can also verify that
\begin{equation} S_n''=\frac{d}{da}\left[\frac{2m^2a}{\omega
_n(a)}\right]=\frac{n^2m}{\omega _n^3}\ll \omega _n^2\end{equation} for
$ma\gg n$ or $ma\ll n$.

Now we consider the backreaction of the excited field modes on the
metric. The divergent sum $\hbar \sum _nS_n$ corresponds to the
infinite zero-point energy of the oscillators $\chi _n$. We assume that
this infinity is absorbed by an appropriate renormalization \cite{PF74}.
Any finite
terms remaining after renormalization will be of order of the squared curvature
$\sim H^4\ll 1$ and we neglect them here (see e.g.~Ref.~\cite{BD78}).
The remaining correction
due to backreaction consists of replacing the effective potential
$V(a)$ by\begin{equation}
V(a)+\Delta V=V(a)-2\hbar \sum _{n=0}^{\infty
}n^3p_n,\label{eq:Vcorr}\end{equation}
where $p_n$ are occupation numbers in the modes $\chi _n$ relative to
an appropriate vacuum and we have inserted the multiplicity factor
$n^3$ [see Ref.~\cite{JVW1}, Eq.~(37) for details].
The average occupation number in the state of the mode $\chi _n$
characterized by $\zeta_n(a)$, relative to a vacuum described by 
$\zeta_n^{(0)}(a)$, is found from Eq.~(\ref{eq:S-Bogolyubov}),
\begin{equation} \left\langle p_n\right\rangle =\left|\beta_n\right| ^2
=\frac{\left|\zeta_n - \zeta_n^{(0)} \right|^2}
{(1-\left|\zeta _n\right|^2)(1-\left|\zeta _n^{(0)}\right|^2)}.
\end{equation} We shall not underestimate $\left\langle p_n\right\rangle $ 
here if we take $\zeta_n^{(0)}\equiv 0$, which corresponds to the vacuum
of the instantaneous diagonalization. The sum over large
$n$ can be estimated using the WKB approximation for $\zeta _n$
[cf.~Eq.~(\ref{eq:zeta-asympt-ans1})] applicable at large $n$ (as shown
in Appendix~\ref{sub:WKB}). The leading term is\begin{equation}
\left|\zeta _n\right|\approx \left|\frac{\dot{\omega }_n}{4\omega
_n^2}\right|.\end{equation} In this limit $\zeta _n$ is small and we can
take $1-\left|\zeta _n\right|^2\approx 1$. At fixed $a$ the squeezing
parameter $\zeta _n(a)$ decays as $n^{-3}$ at large $n$
[Eq.~(\ref{eq:zeta-eq1})]. This decay is fast enough so that the sum
over modes in Eq.~(\ref{eq:Vcorr}) converges. We obtain\begin{eqnarray}
\frac{\Delta V}{V} & \sim  & \frac{1}{V}\sum _{n=0}^{\infty
}\frac{2n^3\left|\zeta _n\right|^2}{1-\left|\zeta _n\right|^2}\nonumber \\
& \approx  & \sum _{n=0}^{\infty }\frac{n^3m^4a^2}{8(n^2+m^2a^2)^3}\approx
\frac{m^2}{32}.
\end{eqnarray}
Since $m\ll 1$ in Planck units, we find that fractional change in
$V(a)$ due to backreaction is small. [The occupation
numbers are computed here in the instantaneous diagonalization vacuum
where $\zeta _n$ is the instantaneous squeezing parameter. The
particle numbers in a higher-order adiabatic vacuum are expected to
be even smaller.]

\subsection{Validity of the Gaussian approximation}

In using the Gaussian ansatz, we disregard the terms which are quartic
in $\chi _n$ but retain the terms quadratic in $\chi _n$. The
Gaussian ansatz gives rise to a quartic term\begin{equation}
\frac{\chi _n^4}{4}\left(\frac{dS_n}{da}\right)^2.\end{equation}
This term can be disregarded at small enough $\chi _n$ if the
following inequality holds,\begin{equation}
\frac{\chi _n^4}{4}\left(\frac{dS_n}{da}\right)^2\ll \chi _n^2\omega
_n^2.\label{eq:WKB-c}\end{equation} The latter condition gives a corridor
around the line $\chi _n=0$ in which the Gaussian ansatz is valid,
$\left|\chi _n\right|<\chi _{\max }$. We can show that the width of this
corridor is never small. The width of the corridor (divided by $a$, since
$\chi _n$ is rescaled by $a$) can be estimated using \begin{equation}
S_n(a)\approx \omega _n+\frac{\dot{\omega }_n}{2\omega
_n}\label{eq:Sn-ans}\end{equation} and Eq.~(\ref{eq:zeta-ans1}). We
find\begin{equation} \left(\frac{\chi _{\max }}{a}\right)^2\sim
\frac{2\omega _n(a)}{dS_n/da}=\frac{2\omega _n\sqrt{V(a)}}{a(S_n^2-\omega
_n^2)}=1+\frac{n^2}{m^2a^2}>1.\end{equation} This expression is bounded
from below and therefore the allowed corridor never shrinks to zero width.

\end{document}